\documentstyle[epsfig]{elsart}
\begin{document}

\makeatletter
\setlength{\clubpenalty}{10000}
\setlength{\widowpenalty}{10000}
\setlength{\displaywidowpenalty}{10000}

\vbadness = 5000
\hbadness = 5000
\tolerance= 1000
\arraycolsep 2pt

\footnotesep 14pt

\if@twoside
\oddsidemargin -17pt \evensidemargin 00pt \marginparwidth 85pt
\else \oddsidemargin 00pt \evensidemargin 00pt
\fi
\topmargin 00pt \headheight 00pt \headsep 00pt
\footheight 12pt \footskip 30pt
\textheight 232mm \textwidth 160mm

\let\@eqnsel = \hfil

\expandafter\ifx\csname mathrm\endcsname\relax\def\mathrm#1{{\rm #1}}\fi
\@ifundefined{mathrm}{\def\mathrm#1{{\rm #1}}}{\relax}

\makeatother

\unitlength1cm
\textheight 233mm

\begin{frontmatter}
\title {{\bf  Left-right symmetry and heavy particle quantum effects.}}

\author{ M. Czakon$^1$},
\author{ J. Gluza$^{1,2}$},
\author{M. Zra\l ek$^1$}

\address{$^1$ Department of Field Theory and Particle Physics, Institute 
of Physics, University of
Silesia, Uniwersytecka 4, PL-40-007 Katowice, Poland}
\address{$^2$ DESY Zeuthen, Platanenallee 6, 15738 Zeuthen, Germany}

\begin{abstract}

We have renormalized a classical left-right model with a bidoublet, and left and right triplets
in the Higgs sector. We focus on oblique corrections and show the interplay between the
top quark, heavy neutrinos and Higgses contribution to the muon $\Delta r$ 
parameter. 
In the SM, custodial symmetry prevents large oblique corrections to 
appear.   
Although in LR models there is no such symmetry
 to make vanish the quadratically diverging terms,
 we have shown, that heavy Higgses contributions to $\Delta r$ are under
control. Also the top contribution to  $\Delta r$, quite different from 
that in the SM,  
is discussed. However, heavy neutrinos  seem to give 
the most important contributions. 
From oblique corrections, they  can be as large as
the SM top one.
Moreover, vertex and box diagrams give additional non-decoupling effects  
and only  concrete numerical estimates  are able to answer whether the 
model is still self-consistent.
\end{abstract}
\end{frontmatter}

\section{Introduction}

Many non-standard models have already been considered in the literature at the quantum level.
They inevitably involved new physical parameters. For instance, if we extend the Higgs sector 
by an additional Higgs doublet,  then mass splitting between  neutral and charged scalars can be
examined \cite{hol1}. 
In the MSSM, supersymmetric particles must also be taken into account and the
analysis is much more sophisticated  \cite{hol2}. This has nothing to do with the gauge sector 
(the gauge group is the same as in the SM), but with the  amount of particles in the game. 
Additional problems appear in models  where heavy neutrinos are introduced.
This has been examined  in the frame of the SM with additional isosinglet neutrino fields 
(which build up Dirac neutrinos) \cite{melo}.
Then, the interplay between heavy neutrinos and  light standard particles on the one hand, and 
nondecoupling
effects \cite{non} on the other, is important            .   
Finally,   non-standard models with extended gauge groups have been considered. 
Let us mention only 
the papers by {\it Senjanovic and Sokorac} 
\cite{senj}, 
where the left-right symmetric  $SU(2)_L \times
SU(2)_R \times U(1)$ has been work out, with conclusion that scalar particle effects 
do not decouple in low-energy processes. This has been done, however, in a model without heavy 
neutrinos\footnote{In the light of new Superkamiokande data \cite{sup}, it is still attractive to 
allow the see-saw mechanism to operate \cite{jez}.} and without renormalization (a class
of diagrams has been chosen, which, after summing, yields a finite answer).
At this point, we should also mention a paper by {\it Soni et al.} \cite{box}, 
where useful limits on the
additional charged gauge boson mass and mixing have been obtained. Note,
that their analysis does not 
require a renormalization procedure (finite box diagrams). A renormalization scheme has been proposed
in \cite{pilo}. It has, however, a different nature from the present, and the consequent analysis
bears no similarity.

In this paper we consider a left-right symmetric model where all of the effects mentioned 
above, come simultaneously into play. 
No numerical estimates will be given. Here we focus only on the renormalization procedure  in
a simple, practical  renormalization framework based on ideas from the SM. 
It can be subsequently used both in  
low- and high-energy physics. To make our presentation clear we 
discuss and give exact relations for oblique corrections. The rest, i.e. influence
of heavy neutrinos and Higgses on vertex and box 
diagrams will be shortly commented, and a detailed analysis will be postponed to \cite{fut}.
As a laboratory we use the muon decay process. 
Let us remind, that the precision experiments, such as the muon decay,  
put extremely stringent constraints on the
oblique corrections \cite{obliq}. These latter, due to the custodial symmetry, 
depend only weakly on the Higgs boson
mass in the SM \cite{cust}. However, there has always
been the danger that the higher order corrections will take the LR model
down. Now, in our case,
there is no symmetry to make vanish the quadratically diverging terms. As the
tree-level phenomenological bounds put the Higgs boson masses in the TeV range, 
there is a real risk that 
it will be impossible to accommodate all of the data, since the radiative corrections will grow 
indefinitely large. The present work shows that the situation is ``reasonable'' and without concrete
fits the model cannot be ruled out.

The organization of the paper is the following. In the next section the most important 
ingredients of the model are described (details can be found in  
\cite{lr,gun1,gun2,glu}).
Next, the renormalization framework will be given and in a subsequent section, a 
quantitative discussion will be 
presented. We end up with conclusions.

\section{The model}

There exists a large class of LR models. They differ by the symmetries
imposed and by the details of the Higgs sector. Much has been written about it in the literature
\cite{gun1,gun2,glu,reszta}, we will
therefore justify our choices only briefly. 

The basic characteristics of the model in question follow from the symmetries. The first is given by
the gauge group, which is:
\begin{equation}
SU(2)_L\otimes SU(2)_R \otimes U(1)_{B-L}.
\end{equation}
It defines the gauge field content, and after the choice of matter fields has been made, 
the interaction of these fields with the gauge bosons and between themselves (to the extent, that further
symmetries may still constrain).

It is by now 
agreed \cite{gun1,gun2} that a minimal model structure should contain these scalar fields:
\begin{enumerate}
\item two Higgs triplets $\Delta_{L,R}$, with quantum numbers $(1,0,2)$ and $(0,1,2)$ respectively,
to generate Majorana neutrino masses through the see-saw mechanism,
\item a Higgs bidoublet $\Phi$, with quantum numbers $(1/2,1/2^*,0)$, to generate charged fermion
masses.
\end{enumerate}
We shall adopt the following convenient representation:
\begin{equation}
\Delta_{L,R} = \left( 
  \begin{array}{cc}
  \delta^+_{L,R}/\sqrt{2} & \delta^{++}_{L,R} \\
  \delta^0_{L,R} & -\delta^+_{L,R}/\sqrt{2} \\
  \end{array} \right), \;\;\;\;
\Phi = \left(
  \begin{array}{cc}
  \phi^0_1 & \phi^+_1 \\
  \phi^-_2 & \phi^0_2 \\
  \end{array} \right).
\end{equation}

Even after the choice of the fields, the allowed
lagrangians will still have much freedom left. Additional constraints follow from the
left-right symmetry:
\begin{equation}
W_L \leftrightarrow W_R, \;\;\; \Psi_L \leftrightarrow \Psi_R, \;\;\; \Delta_L \leftrightarrow \Delta_R,
\;\;\; \Phi \leftrightarrow \Phi^\dagger,
\end{equation}
where $W_{L,R}$ are gauge fields associated to the left and right $SU(2)$ gauge groups, and $\Psi_{L,R}$ are
left and right fermion fields. Imposing
this symmetry leads not only to several simplifications, but also to a restoration of parity invariance at
high energies (in the unbroken phase). This should be considered the most important argument for its 
introduction.

The most general Higgs potential  allowed by the adopted symmetries, that we consider in this paper, 
was discussed in  \cite{gun1,gun2,glu}.

The scalar fields can develop vacuum expectation values through the spontaneous symmetry breaking mechanism:
\begin{equation}
<\Delta_{L,R}> = \left( 
  \begin{array}{cc}
  0 & 0 \\
  v_{L,R}/\sqrt{2} & 0 \\
  \end{array} \right), \;\;\;\;
<\Phi> = \left(
  \begin{array}{cc}
  \kappa_1/\sqrt{2} & 0 \\
  0 & \kappa_2/\sqrt{2} \\
  \end{array} \right).
\end{equation}
A very careful analysis of the symmetry breaking pattern has been given by {\it Deshpande et al.}
\cite{gun2}.
Surprisingly, it turns out that due to see-saw type relations between the vacuum expectation values
and the coupling constants, the possible values of the parameters are strongly constrained, to the
point that avoiding fine tunings requires setting several of them to zero. Therefore, as it has been
argued, the most convenient results can be obtained with
$v_L = 0$ and $\beta$ terms in the Higgs potential put to zero 
(see \cite{gun2}).  

Several other approximations are in order. 
Since we do not consider CP violation effects, we assume all of the Higgs potential parameters to be real,
which yields real VEVs.
Furthermore, strong suppression of $FCNC$, requires that either of $\kappa_{1,2}$ be very small, or 
vanishing \cite{gun2,fcnc}. 
To simplify the expressions, the one-loop analysis will be done with
$\kappa_2 = 0$\footnote{due to the symmetry of the Higgs potential, the model is symmetric with respect
to replacement $\kappa_1 \leftrightarrow \kappa_2$.}. The general character
of the radiative corrections will not be changed by this last assumption. 

There are
twenty real fields at our disposal. Since some of them are charged, there should be fourteen distinct
fields (unconnected by symmetries). This number is further reduced, as some of them will become 
Goldstone bosons for the gauge fields (their number is fixed by the symmetry breaking pattern, to be
four). Thus, we end up with ten physical fields. Let us now write them in terms of the original 
scalars.

We begin with four neutral scalars $H^0_a$, $a=0,1,2,3$, and two neutral pseudo-scalars $A^0_{1,2}$\footnote{
the lack of mixing between the physical scalars is only approximate and follows from the large difference
of scales between $\kappa_1$ and $v_R$}:
\begin{eqnarray}
\phi^0_1 &=& \frac{1}{\sqrt{2}}(H^0_0+i\phi^{0i}_1), \\
\phi^0_2 &=& \frac{1}{\sqrt{2}}(H^0_1+iA^0_1), \\
\delta^0_R &=& \frac{1}{\sqrt{2}}(H^0_2+i\delta^{0i}_R), \\
\delta^0_L &=& \frac{1}{\sqrt{2}}(H^0_3+iA^0_2),
\end{eqnarray}
where $\delta^{0i}_R$ and $\phi^{0i}_1$, are massless and thus functions of the Goldstone fields:
\begin{equation}
\left( \begin{array}{c} \delta^{0i}_R \\ \phi^{0i}_1 \end{array} \right) = 
\left( \begin{array}{cc} -s_g^n & -c^n_g \\ c_g^n & -s^n_g \end{array} \right)
\left( \begin{array}{c} G^0_1 \\ G^0_2 \end{array} \right),
\end{equation}
with (see next section for definitions of the gauge sector mixing parameters):

\begin{eqnarray}
s^n_g & = & -\frac{gv_R}{M_{Z_1}}\frac{s}{c_M} = \frac{g\kappa}{2M_{Z_2}}(c_Mc-s/c_W), \\
c^n_g & = & \frac{gv_R}{M_{Z_2}}\frac{c}{c_M} = \frac{g\kappa}{2M_{Z_1}}(c/c_W+c_Ms).
\end{eqnarray}

The charged sector contains two singly charged Higgses $H^\pm_{1,2}$, two singly charged Goldstone bosons
$G^\pm_{1,2}$, and two (physical) doubly charged Higgses $\delta^{\pm \pm}_{L,R}$:
\begin{eqnarray}
\delta_L^\pm &=& H_1^\pm, \\
\delta_R^\pm &=& s_g^c H_2^\pm\pm i c_g^c G^\pm_2, \\
\phi_1^\pm &=& c_g^c H_2^\pm\mp i s_g^c G^\pm_2, \\
\phi_2^\pm &=& \mp iG^\pm_1,
\end{eqnarray}
with:
\begin{equation}
c_g^c = \frac{\sqrt{2}v_R}{\sqrt{\kappa_1^2+2v_R^2}}, \;\;\;\;
s_g^c = \frac{\kappa_1}{\sqrt{\kappa_1^2+2v_R^2}}.
\end{equation}
By orthogonality of the respective mixing matrices, $s^n_g$, $s^c_g$, and $c^n_g$, $c^c_g$, define sines and
cosines of some mixing angles.

We will not  specify details of the gauge-Higgs and Yukawa interactions. They can be 
found in \cite{glu}.
The charged gauge-lepton current is:

\begin{equation}
{\cal L}_{CC} = \frac{g}{\sqrt{2}}\left(\overline{N}\gamma^\mu K_L P_L l W^+_{1\mu}
+\overline{N}\gamma^\mu K_R P_R l W^+_{2\mu} \right)+h.c.\;\;.
\end{equation}

Here $K_L,K_R$ are neutrino mixing matrices. It is  
justified by present  experimental data and neutrino mass generation mechanisms, 
that light neutrinos couple strongly to leptons
through left currents, and heavy neutrinos through right currents. Even more than that, 
since in all
considered processes light neutrino masses can be neglected, we may assume that the interaction between
leptons and light neutrino states is diagonal. The most important consequence is that one
loop corrections to muon decay are inserted to the tree level $W_1$ diagrams 
(not to the $W_2$ diagrams, and $W_1-W_2$ mixing is not considered).  

The neutral current is \cite{glu1}:
\begin{eqnarray}
{\cal L}_{NC}&=& \frac{e}{2s_Wc_W}\left( \overline{l}\gamma^\mu(A^{1l}_L P_L+A^{1l}_R P_R
)lZ_{1\mu}+\overline{N}\gamma^\mu
(A^{1\nu}_L \Omega_L P_L+A^{1\nu}_R \Omega_R P_R)NZ_{1\mu} \right. \nonumber
\\
& &\;\;\;\;\;\;\;\;\;\;+ \left. \overline{l}\gamma^\mu(A^{2l}_L P_L+A^{2l}_R P_R
)lZ_{2\mu}+\overline{N}\gamma^\mu
(A^{2\nu}_L \Omega_L P_L+A^{2\nu}_R \Omega_R P_R)NZ_{2\mu} \right) \nonumber
\\
& & \;\;\;\;\;\;\;\;\;\; -e\left(\overline{l}\gamma^\mu lA_{\mu} \right),
\end{eqnarray}
where:
\begin{equation}
\Omega_L = K_L K_L^\dagger,\;\;\;\; \Omega_R = K_R K_R^\dagger,
\end{equation}
Here also, we can state that $\Omega_L$ couples diagonally light to light neutrino states, whereas 
$\Omega_R$ couples (but in general non-diagonally) mostly heavy states. Couplings between light and heavy states
can be neglected in most but a few specific cases where the diagram is proportional to heavy neutrino
mass squared and must be treated separately.

\section{Renormalization}

From the point
of view of the SM, precision tests based on four-fermion reactions may be considered complete, if
we forget about a few recompilations of results. The case is much different for the LR model. Since 
the analysis is simpler, when there is no gauge boson final states, we 
shall begin here a 
systematic work on radiative corrections in this specific situation. 
Thus, the only wave functions that need
to be renormalized are fermionic (plus the photonic one,
as we shall see in a 
moment). Let us remind, that the wave function renormalization constants
serve only the purpose of 
properly normalizing the amplitudes.
The real choice comes with the physical input parameters. Forget for a moment Higgs and Yukawa 
sectors  and focus on the gauge one.
The free parameters are:
\begin{equation}
g,\;\;\;g',\;\;\;\kappa_1,\;\;\;\kappa_2,\;\;\;v_R.
\end{equation}
All physical parameters (mixing angles and masses) can be expressed in
their terms, namely
gauge boson mixing  matrices defined as follows
($\kappa_+ = \sqrt{\kappa_1^2+\kappa_2^2}$):
\begin{eqnarray}
\left( \begin{array}{c} W_L^\pm \\ W_R^\pm \end{array} \right) & = &
\left( \begin{array}{cc} \cos{\zeta} & \sin{\zeta} \\ -\sin{\zeta} &
\cos{\zeta} \end{array} \right)
\left( \begin{array}{c} W^\pm_1 \\ W^\pm_2 \end{array} \right),
\\ \nonumber \\ \nonumber
\left( \begin{array}{cc} W_L^3 \\ W_R^3 \\ B \end{array} \right) & = &
\left( \begin{array}{ccc} c_Wc & c_Ws & s_W \\ -s_Ws_Mc-c_Ms &
-s_Ws_Ms+c_Mc
& c_Ws_M \\   
                         -s_Wc_Mc+s_Ms & -s_Wc_Ms -s_Mc & c_Wc_M
\end{array}
\right)
\left( \begin{array}{cc} Z_1 \\ Z_2 \\ A \end{array} \right), \\
\end{eqnarray}

where:
\begin{eqnarray}
c_W \equiv \cos{\theta_W}, \;\;\; s_W &\equiv& \sin{\theta_W}, \;\;\; c_M
\equiv \frac{\sqrt{\cos{2\theta_W}}}{\cos{\theta_W}},
             \;\;\; s_M \equiv \tan{\theta_W} \nonumber \\
s &\equiv& \sin{\phi}, \;\;\; c \equiv \cos{\phi}.
\end{eqnarray}

gives
\begin{equation}
\tan{2\zeta} = -\frac{2\kappa_1\kappa_2}{v_R^2}, \;\;\;\; \sin{2\phi} =
-\frac{g^2\kappa^2_+\sqrt{\cos{2\theta_W}}}{2\cos^2{\theta_W}
(M_{Z_2}^2-M_{Z_1}^2)}.
\end{equation}

The angle $\theta_W$, which we call in analogy to SM, the Weinberg angle,
is connected to the electric charge
and the coupling constants $g$, $g'$:
\begin{equation}
g=\frac{e}{\sin{\theta_W}}, \;\;\;\; g'=\frac{e}{\sqrt{\cos{2\theta_W}}}.
\end{equation}

Similarly, 
the masses of the gauge bosons are given by the following equations:

\begin{eqnarray}
M^2_{W_{1,2}} &=& \frac{g^2}{4}\left(\kappa_+^2+v_R^2\mp\sqrt{v_R^4+
4\kappa_1^2\kappa^2_2}\right), \\
M^2_{Z_{1,2}} &=& \frac{1}{4}\left(\left((g^2\kappa_+^2+2v_R^2(g^2+g'^2)
\right)\right. \\ \nonumber
              & &  \mp\left.\sqrt{(g^2\kappa_+^2+2v_R^2(g^2+g'^2))^2-
4g^2(g^2+2g'^2)\kappa^2_+v_R^2}\right).
\end{eqnarray}

So, finally, we can use in the renormalization procedure 
the following set of physical parameters 
\begin{equation}
e,\;\;\;M_{W_1},\;\;\;M_{W_2},\;\;\;M_{Z_1},\;\;\;M_{Z_2}.
\end{equation}

The electromagnetic
coupling constant and the light boson masses are known and natural, 
just the same way as they are
in the SM. Whether to use, in numerical analyses, $M_{W_1}$ or take it 
from the muon decay remains an
issue reserved for future. 

Let us recapitulate the main points of the scheme:
\begin{enumerate}
\item we only renormalize the fermion wave-functions, no gauge boson renormalization constant is introduced,
except for the photon (and this only to define the electric charge),
\item the masses are renormalized on-shell ({\it i.e} the poles of the respective propagators are fixed
at the physical masses),
\item the mixing angles of the gauge boson sector are renormalized using their relation to the gauge boson
masses.
\end{enumerate}

We now have to make a remark about the self-consistency of our scheme when
we assume that $\kappa_2 = 0$. There is then no mixing between the
charged gauge bosons at tree-level (see Eq.(24)). 
A divergent contribution to this mixing shows up at one-loop level 
through fermion loops. Thus, although we can safely keep $\kappa_2 = 0$, a counter-term will still
be necessary. However, in muon decay we do not have to bother about it,
since tree
level diagrams with $W_2$ 
transitions are negligible \cite{jap}. 

We now return to the Higgs sector. We assumed that certain coupling constants were zero, without
any symmetries imposed. This would lead to nonrenormalizability. If we were to consider the 
renormalization of this sector, additional counter-terms would have to 
be introduced. Fortunately, 
for the muon decay case this will not be necessary.

The scheme
described above is an extension of the work of {\it Sirlin} \cite{eight}. 

\subsection{Fermion Propagator Renormalization}

\begin{figure}
\begin{center}
\epsfig{file=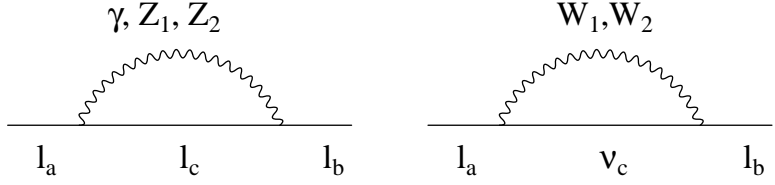}
\caption{Diagrams entering the charged lepton self-energy, with neglect of heavy neutrino contributions}
\label{fermion}
\end{center}
\end{figure}

Let us turn to the fermion propagator renormalization. If we introduce the notation $-i\Sigma_{ba}$ for
the irreducible contribution to the transition from $a$ to $b$, and decompose it according to the Lorentz
structure:
\begin{equation}
\Sigma_{ba}(p) = \hat{p} P_L\Sigma^{\gamma L}_{ba}(p^2)
+\hat{p} P_R\Sigma^{\gamma R}_{ba}(p^2)
+P_L\Sigma^{1 L}_{ba}(p^2)
+P_R\Sigma^{1 R}_{ba}(p^2),
\end{equation}
then the following set of renormalization conditions can be given (the hat denotes renormalized quantities):
\begin{eqnarray}
m^l_a\hat{\Sigma}^{\gamma L}_{ba}(m^{l\;2}_a)+\hat{\Sigma}^{1R}_{ba}(m^{l\;2}_a) &=& 0, \nonumber\\ & & \nonumber \\
m^l_a\hat{\Sigma}^{\gamma R}_{ba}(m^{l\;2}_a)+\hat{\Sigma}^{1L}_{ba}(m^{l\;2}_a) &=& 0, \nonumber\\& & \\
m^l_a\hat{\Sigma}^{\gamma L}_{ab}(m^{l\;2}_a)+\hat{\Sigma}^{1L}_{ab}(m^{l\;2}_a) &=& 0, \nonumber\\& & \nonumber \\
m^l_a\hat{\Sigma}^{\gamma R}_{ab}(m^{l\;2}_a)+\hat{\Sigma}^{1R}_{ab}(m^{l\;2}_a) &=& 0, \nonumber
\end{eqnarray}
for $b\neq a$, and:
\begin{equation}
\begin{array}{l}
\hat{\Sigma}^{\gamma L}_{aa}(m^{l\;2}_a) = \hat{\Sigma}^{\gamma R}_{aa}(m^{l\;2}_a), \\ \\
\hat{\Sigma}^{\gamma L}_{aa}(m^{l\;2}_a)+\hat{\Sigma}^{\gamma R}_{aa}(m^{l\;2}_a)
+2m^{l\;2}_a\left(\hat{\Sigma}^{\gamma L\;'}_{aa}(m^{l\;2}_a)+\hat{\Sigma}^{\gamma R\;'}_{aa}(m^{l\;2}_a)\right) \\ \\
+2m^{l}_a \left( \hat{\Sigma}^{1L\;'}_{aa}(m^{l\;2}_a)+\hat{\Sigma}^{1R\;'}_{aa}(m^{l\;2}_a) \right) = 0,
\end{array}
\end{equation}
for the diagonal case. These equations can be satisfied by introducing mass counter-terms and 
matrices
of left(right) wave function constants. The situation can be simplified if the transition occurs through
light particles. In the following, we will neglect the heavy neutrino contributions\footnote{not to break
renormalizability, we have to include all the diagrams. Nevertheless, we can simplify the finite parts of
the radiative corrections, by assuming that all neutrinos are massless. The difference will be accounted for in numerics \cite{fut}.}. 
Thus, all the self-energies are diagonal (as already discussed, $K_L$ neutrino mixing matrix 
connected with light neutrinos is assumed to be diagonal). 
We now need only diagonal wave function renormalization constants. They can be
expressed through (see \cite{hol1}, for the definition of $\Delta$):
\begin{eqnarray}
\delta Z^{l_a}_L &=& 
\Sigma^{\gamma L}_{aa}(m^{l\;2}_a)
+m^{l\;2}_a\left(\Sigma^{\gamma L\;'}_{aa}(m^{l\;2}_a)+\Sigma^{\gamma R\;'}_{aa}(m^{l\;2}_a)\right) \nonumber \\
& & +m^{l}_a \left( \Sigma^{1L\;'}_{aa}(m^{l\;2}_a)+\Sigma^{1R\;'}_{aa}
(m^{l\;2}_a) \right), \\ \nonumber \\
\delta Z^{l_a}_R &=& 
\Sigma^{\gamma R}_{aa}(m^{l\;2}_a)
+m^{l\;2}_a\left(\Sigma^{\gamma L\;'}_{aa}(m^{l\;2}_a)+\Sigma^{\gamma R\;'}_{aa}(m^{l\;2}_a)\right) \nonumber \\
& & +m^{l}_a \left( \Sigma^{1L\;'}_{aa}(m^{l\;2}_a)+\Sigma^{1R\;'}_{aa}
(m^{l\;2}_a) \right).
\end{eqnarray}
Note that terms proportional to the masses are non-negligible only for the photonic transition\footnote{
since we consider only light particles, these terms are proportional to $\frac{(m^l_a)^2}{M^2}$, where
$M$ is a heavy boson mass. This can, obviously, be neglected.}.
The diagrams that enter the calculation are depicted on fig.~\ref{fermion}.
The left-right symmetry has for consequence the equality of the divergent parts of $\delta Z^{l_a}_L$ and
$\delta Z^{l_a}_R$. We get the following simple result:
\begin{eqnarray}
   \delta Z^{l_a}_L &=& \frac{\alpha}{4\pi}\left(-\Delta(m^l_a)
       -\frac{1}{2s_W^2}\Delta(M_{W_1})
       -\frac{(A^{l1}_L)^2}{4s_W^2c_W^2}\Delta(M_{Z_1})
       -\frac{(A^{l2}_L)^2}{4s_W^2c_W^2}\Delta(M_{Z_2}) \right. \nonumber \\
    & & \left. -2\log{\frac{\lambda^2}{m^{l\;2}_a}}- 4 + \frac{(A^{l1}_L)^2}{8s_W^2c_W^2}
          +\frac{(A^{l2}_L)^2}{8s_W^2c_W^2} +\frac{1}{4s_W^2}\right), \\ \nonumber \\
   \delta Z^{l_a}_R &=& \frac{\alpha}{4\pi}\left(-\Delta(m^l_a)
       -\frac{1}{2s_W^2}\Delta(M_{W_2})
       -\frac{(A^{l1}_R)^2}{4s_W^2c_W^2}\Delta(M_{Z_1})
       -\frac{(A^{l2}_R)^2}{4s_W^2c_W^2}\Delta(M_{Z_2}) \right. \nonumber \\
    & & \left. -2\log{\frac{\lambda^2}{m^{l\;2}_a}}- 4 + \frac{(A^{l1}_R)^2}{8s_W^2c_W^2}
          +\frac{(A^{l2}_R)^2}{8s_W^2c_W^2} +\frac{1}{4s_W^2}\right).
\end{eqnarray}
The above do not depend on the lepton species, apart from the infrared logarithms, and even these
will cancel in the photon vertex due to the electromagnetic Ward identity. The neutrino
constants are quite similar. We do not give the diagrams, since they are analogous. The result is:
\begin{eqnarray}
   \delta Z^{\nu_a}_L &=& \delta Z^{\nu_a}_R = \frac{\alpha}{4\pi}\left(
       -\frac{1}{2s_W^2}\Delta(M_{W_1})
       -\frac{(A^{\nu 1}_L)^2}{4s_W^2c_W^2}\Delta(M_{Z_1})
       -\frac{(A^{\nu 2}_L)^2}{4s_W^2c_W^2}\Delta(M_{Z_2}) \right. \nonumber \\
    & & \;\;\;\;\;\;\;\;\;\;\;\;\;\;\;\;\;\;\; \left. +\frac{(A^{\nu 1}_L)^2}{8s_W^2c_W^2}
          +\frac{(A^{\nu 2}_L)^2}{8s_W^2c_W^2} +\frac{1}{4s_W^2}\right),
\end{eqnarray}
Let us stress, that these constants are equal to one another due to the Majorana nature of the
neutrinos.

If we were to have non-diagonal transitions, then renormalization of mixing matrices would be
required as pointed out first by {\it Denner and Sack} \cite{sack}, 
and later for the case of neutrinos by
{\it Kniehl and Pilaftsis} \cite{kniehl}. 
Thanks to our simplifications, we do not have to bother about it right now.

\subsection{The Photon Propagator}

Although in four-fermion processes we do not have to renormalize external photon lines, this is necessary
to define the electric charge counter-term. 
Let us give the photon one-loop self-energy 
contribution\footnote{$i\Pi_{ba}$ is the transverse part of the irreducible 
contribution to the transition from gauge boson $a$ to $b$, see
\cite{hol} 
for the definition of the $F$ function.}:
\begin{eqnarray}
\Pi_{\gamma\gamma}(p^2) &=& \frac{\alpha}{4\pi}\left(
-\frac{4}{3}\sum_{ferm.} Q_f^2 \left(\Delta(m_f)p^2 +(2m_f^2+p^2)F(p,m_f,m_f)
-\frac{p^2}{3}\right)\right.\nonumber \\
& & \;\;\;\;\;\;\;\;\; +3\sum_{i=1,2}\left(\Delta(M_{W_i})p^2
+(p^2+\frac{4}{3}M_{W_i}^2)F(p,M_{W_i},M_{W_i}) \right)
\nonumber \\ 
& &\;\;\;\;\;\;\;\;\left.-\frac{1}{3} \sum_{Higgs}Q^2_H\left(\Delta(M_H)p^2
+(p^2-4M_H^2)F(p,M_H,M_H)+\frac{2}{3}p^2\right)\right). \nonumber \\ \label{photo1}
\end{eqnarray}
One can check readily that this vanishes at zero momentum, which means that the photon remains 
massless.
The derivative of this expression is the photon wave-function renormalization constant:
\begin{eqnarray}
\delta Z^\gamma \equiv \Pi'_{\gamma\gamma}(0) &=& \frac{\alpha}{4\pi} \left(-\frac{4}{3} \sum_{ferm.}
Q^2_f \left(\Delta(m_f)\right)\right. 
+3\sum_{i=1,2}\left(\Delta(M_{W_i})+\frac{2}{9}\right) \nonumber \\
& & \;\;\;\;\;\;\;\left. -\frac{1}{3}\sum_{Higgs}Q_H^2\left(\Delta(M_H)\right)\right). \label{photo2}
\end{eqnarray}
This should be compared to the SM result. We see that the fermions gave the same contribution, which
was expected since the electromagnetic interaction is the same in all models. The term coming from
the additional charged gauge boson is the same as the one for $W_1$, apart from the mass difference.
The new addition is the term coming from charged Higgs loops. 
Its contribution is not numerically large due to logarithmic nature.

\subsection{The Electric Charge Counter-Term}

The electric charge is defined through the Thomson scattering amplitude. In 
practice it is measured in different processes, like the Hall or Josephson effect, but the former
definition is closer to our perturbative methods. Thanks to the electromagnetic Ward identity, $\delta e$
will be infrared divergence free.

\begin{figure}
\begin{center}
\epsfig{file=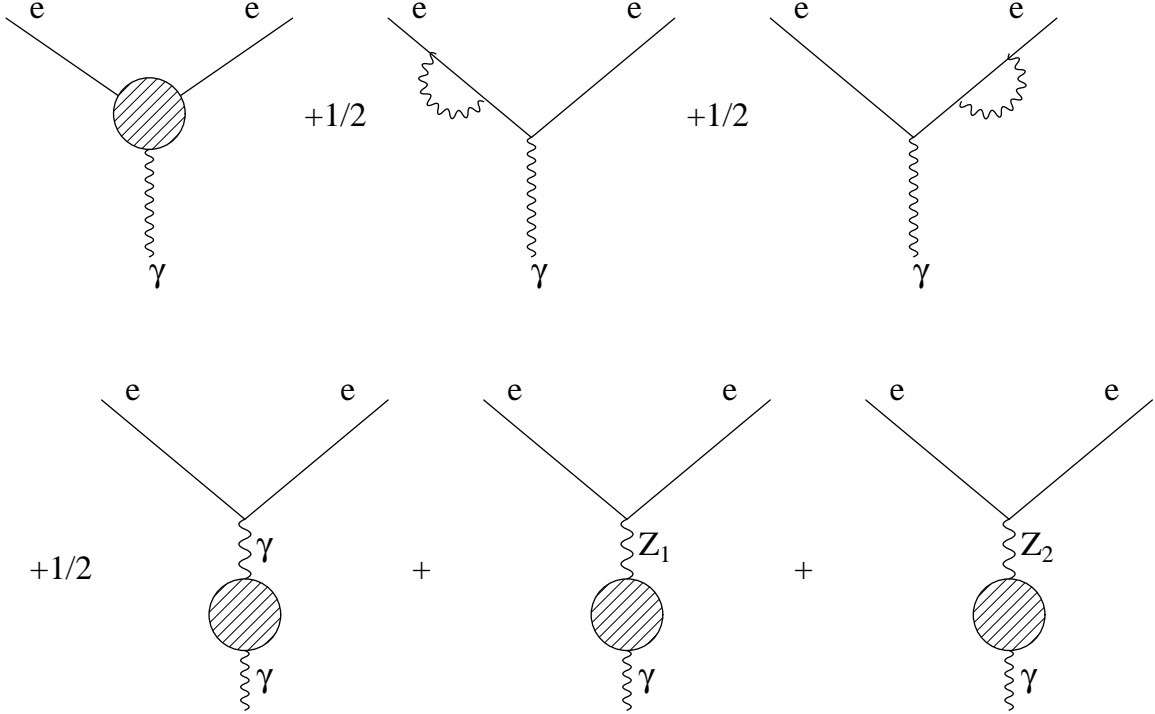}
\end{center}
\caption{Diagrams entering the electric charge counter-term.}
\label{deltae}
\end{figure}

The one-loop contribution to the $\gamma ll$ amplitude, may only have a divergent vector part, otherwise it 
would not 
be multiplicatively renormalizable. It is thus an important check on the calculations to verify the vanishing
of the divergence of the axial part. The diagrams entering this contribution are depicted on 
fig~\ref{deltae}.
Again, the left-right symmetry of the model causes the vanishing of the divergence of the axial part
of the photon vertex corrections and the equality of those of the left and right lepton renormalization 
constants.
Thus, what remains is a constraint on the $\gamma Z_1$ and $\gamma Z_2$ transitions at zero momentum:
\begin{equation}
\label{constr1}
\frac{1}{M_{Z_1}}\left(A^{l1}_L-A^{l1}_R\right)\Pi_{\gamma Z_1}(0) =
\frac{1}{M_{Z_2}}\left(A^{l2}_L-A^{l2}_R\right)\Pi_{\gamma Z_2}(0).
\end{equation}
Now, these transitions read:
\begin{eqnarray}
\Pi_{\gamma Z_1}(0) &=& \frac{\alpha}{4\pi}\frac{2}{s_W^2}\left(\Delta(M_{W_1})
(c/c_W+c_Ms)M_{W_1}^2 \right. \\ & & \;\;\;\;\;\;\;\;\;
+\left. \Delta(M_{W_2})(\sqrt{2}s_g^n c^c_g-c^n_g s^c_g)M_{W_2}M_{Z_1}
\right), \nonumber \\ & & \nonumber \\
\Pi_{\gamma Z_2}(0) &=& \frac{\alpha}{4\pi}\frac{2}{s_W^2}\left(\Delta(M_{W_1})
(s/c_W-c_Mc)M_{W_1}^2 \right. \\ & & \;\;\;\;\;\;\;\;\;
+\left. \Delta(M_{W_2})(\sqrt{2}c_g^n c^c_g+s^n_g s^c_g)M_{W_2}M_{Z_2}
\right). \nonumber
\end{eqnarray}
The eq.~\ref{constr1} is satisfied, as we see. Since everything seems to be correct, we can now calculate
the electric charge counter-term. Let us remind that from fig.~\ref{deltae}, follows that\footnote{
$-ie\Lambda_{\gamma ll}(0)$, is the one-loop correction to the $\gamma ll$ vertex at zero momentum transfer, 
where $l$ stands as usual for the lepton.}:
\begin{eqnarray}
\frac{\delta e}{e} &=& -\left( \Lambda_{\gamma ll}(0)+\frac{1}{2}(\delta Z^l_L+\delta Z^l_R) 
+\frac{1}{2}\delta Z^\gamma \right. \\
& & +\frac{1}{4s_Wc_W}\left.\left(\frac{1}{M_{Z_1}}\left(A^{l1}_L
+A^{l1}_R\right)\Pi_{\gamma Z_1}(0)+\frac{1}{M_{Z_2}}\left(A^{l2}_L
+A^{l2}_R\right)\Pi_{\gamma Z_2}(0) \right)\right)\nonumber
\end{eqnarray}
Evaluating this expression leads to:
\begin{eqnarray}
\frac{\delta e}{e} &=& \frac{\alpha}{4\pi}\left(\frac{2}{3}\sum_{ferm.}
Q_f^2\left(\Delta(m_f)\right)+\sum_{i=1,2}\left(-\frac{7}{2}\Delta(M_{W_i})
-\frac{1}{3}\right)+\frac{1}{6}\sum_{Higgs}Q_H^2\left(\Delta(M_H)\right)\right).
\nonumber \\
\end{eqnarray}
The non-abelian couplings from both of the $SU(2)$ groups lead to a modification of the {\it QED}
Ward identity:
\begin{equation}
\frac{\delta e}{e} = -\frac{1}{2}\delta Z^\gamma-\frac{\alpha}{4\pi}2\sum_{i=1,2}
\Delta(M_{W_i})
\end{equation}
A similar result in the SM has been easily justified with use of an abelian Ward identity and a few
other constraints. An analogous proof would be more difficult in this case.

Let us note at last, that due to a coincidence between the number of fermion generations and charge
assignments, and the number of charged Higgses $\delta e$ turns out to be finite. This was first
noted by {\it Duka} \cite{duka}\footnote{This is an interesting observation, since it has consequences
for the running of the electromagnetic coupling constant. Remark, that at some energy scale larger than
the largest mass of the charged particles, $\alpha$ will cease running. This means that we will have
problems embedding this left-right model in a GUT. Note that
some parameters of the Higgs potential have been made vanishing by hand, and it has been argued that
this may find its reason in some GUT model.}.

\subsection{The Weinberg Angle Counter-Term}

The most difficult part of the renormalization scheme comes now. The approximations that $\kappa_2=0$, 
together
with the remarks at the beginning of the present section, 
allow us to forget about the $\zeta$ angle. 
The $\phi$
angle counter-term is needed only in neutral current processes. What remains, is the Weinberg angle.
In principle, it can be expressed analytically through the gauge boson masses (but this leads to
a fourth order equation). When $\kappa_2 = 0$, the
appropriate expression is rather simple. It should not be
used, however, to derive a counter-term, since the result would not take into account the counter-term to 
$\kappa_2$. The situation is fortunately simpler because $\delta s_W^2$ can be obtained from a system of
linear equations.

Let us start by writing the following (see Eqs.(26,27)):
\begin{eqnarray}
M_{W_2}^2+M_{W_1}^2 &=& \frac{1}{2} g^2 (\kappa_+^2+v_R^2), \\ & & \nonumber \\
M_{Z_2}^2+M_{Z_1}^2 &=& \frac{1}{2} (g^2 \kappa_+^2+2v_R^2(g^2+g'^2)), \\ & & \nonumber \\
M_{Z_2} M_{Z_1} &=& \frac{1}{2} g^2 \sqrt{1+2 \frac{g'^2}{g^2}} \kappa_+ v_R.
\end{eqnarray}
To incorporate one-loop corrections, we have to expand these equations to first order:
\begin{eqnarray}
\delta M_{W_2}^2+\delta M_{W_1}^2 &=& \frac{1}{2} \delta g^2 (\kappa_+^2+v_R^2)
+\frac{1}{2} g^2 (\delta \kappa_+^2+\delta v_R^2), \\ & & \nonumber \\
\delta M_{Z_2}^2+\delta M_{Z_1}^2 &=& \frac{1}{2} \delta g^2 \kappa_+^2+v_R^2(\delta g^2+\delta g'^2)
+\frac{1}{2} g^2 \delta \kappa_+^2+\delta v_R^2(g^2+g'^2), \nonumber \\ & &  \\
M_{Z_2} \delta M_{Z_1} +M_{Z_1} \delta M_{Z_2} &=& \frac{1}{2} \delta\left( g^2 \sqrt{1+2 \frac{g'^2}{g^2}} 
\right)\kappa_+ v_R \nonumber \\
&+& \frac{1}{4}g^2 \sqrt{1+2 \frac{g'^2}{g^2}}\left(\frac{v_R}{\kappa_+}\delta \kappa_+^2
+\frac{\kappa_+}{v_R} \delta v_R^2 \right), \nonumber \\
\end{eqnarray}
where:
\begin{equation}
\delta\left( g^2 \sqrt{1+2 \frac{g'^2}{g^2}} \right) = \frac{1}{g^2 \sqrt{1+2 g'^2/g^2}}\left(
(g^2+g'^2)\delta g^2+g^2 \delta g'^2 \right).
\end{equation}
If $\delta \kappa_+^2 = \delta v_R^2 = 0$, then one could even take $\delta s_W^2$ from any of these 
equations. The fact is however, that we would have to include tadpole
diagrams in gauge boson self-energies. This would be considerably more difficult, since it would 
have introduced
the Higgs sector. We thus silently put the tadpoles to zero through appropriate renormalization constants,
but use all of the equations to eliminate $\delta \kappa_+^2$ and $\delta v_R^2$. 
Now, since the coupling constants are related to the Weinberg angle through:
\begin{equation}
g=\frac{e}{\sin{\theta_W}}, \;\;\;\; g'=\frac{e}{\sqrt{\cos{2\theta_W}}},
\end{equation}
the following result ensues:
\begin{eqnarray}
\delta s_W^2 &=& 2 c_W^2 \frac{(\delta M_{Z_2}^2+\delta M_{Z_1}^2)-(\delta M_{W_2}^2+\delta M_{W_1}^2)}
{(M_{Z_2}^2+M_{Z_1}^2)-(M_{W_2}^2+M_{W_1}^2)} \nonumber \\ & & \nonumber \\
& & +\frac{1}{2} \frac{(M_{W_2}^2+M_{W_1}^2)(\delta M_{Z_2}^2+\delta M_{Z_1}^2)
+(M_{Z_2}^2+M_{Z_1}^2)(\delta M_{W_2}^2+\delta M_{W_1}^2)}
{\left((M_{Z_2}^2+M_{Z_1}^2)-(M_{W_2}^2+M_{W_1}^2)\right)^2} \nonumber \\ & & \nonumber \\
& & -\frac{1}{2} \frac{(2 M_{Z_1}^2+M_{Z_2}^2)\delta M_{Z_1}^2+(2 M_{Z_2}^2+M_{Z_1}^2)\delta M_{Z_2}^2}
{\left((M_{Z_2}^2+M_{Z_1}^2)-(M_{W_2}^2+M_{W_1}^2)\right)^2}. 
\end{eqnarray}
Although it looks symmetric, this is also much more complicated than in the SM. The last important point
is this equation:
\begin{equation}
\left((M_{Z_2}^2+M_{Z_1}^2)-(M_{W_2}^2+M_{W_1}^2)\right) = \frac{g^2}{2c_M^2 c_W^2} v_R^2.
\end{equation}
It sets the scale of the radiative corrections, which will be the subject 
of study of the next section.

\subsection{Methods and Cross-Checks}

We have presented the necessary parts of the renormalization scheme. When applying the formulae,
and most of all the $\delta s_W^2$ constant we are facing the problem of dealing with enormous numbers
of diagrams. To make this study feasible, certain computer methods have been developed. We would like 
to describe them in some detail. 

Most of the calculations have been performed with the program FORM \cite{verm}. 
This includes the basic one-loop
diagrams with undefined coupling constants. They have been tested on the Standard Model. All of the results
given in standard works on radiative corrections to the SM have been recovered. The vertex and fermion
self-energy corrections have been coded by analogy to the SM, and it has been verified that they lead
to the same results in appropriate limits. It has been thus assumed, that the formulae are correct, and
that the only source of problems would be the couplings, statistical factors and signs in the gauge
boson self-energies. Because of the
large Higgs sector, a special program has been written in C, to parse lagrangians and produce input for FORM
in terms of the previously tested functions. It again, has been able to recover SM results. To verify
the correctness of the results for the full LR model, we used a set of constraints on the divergent parts
of the self-energies following from the relation between its broken and its unbroken phases\footnote{
the divergent parts of the gauge boson wave function renormalization constants can be found without any
explicit assumptions on boson renormalization equations. They are simply the parts of the self-energies
proportional to the momentum squared.}\footnote{
To simplify notation we  identify the neutral sector
mixing matrix given in Eq.(22) with the following matrix   
\begin{equation}
\left( \begin{array}{ccc} x_1 & x_2 & x_3 \\ y_1 & y_2 & y_3 \\ v_1 & v_2
&
v_3 \end{array} \right).
\end{equation}
}:
\begin{eqnarray}
\left( \delta Z^{W_2} \right)_{div.} & = & \left( \delta Z^{W_1} \right)_{div.}, \\ & & \nonumber \\
v_1^2 \left(\delta Z^{Z_2}-(x_2^2+y_2^2) \delta Z^{W_2} \right)_{div.} & = &
v_2^2 \left(\delta Z^{Z_1}-(x_1^2+y_1^2) \delta Z^{W_1} \right)_{div.}, 
\end{eqnarray}
and
\begin{eqnarray}
\left(\delta Z^{Z_1} \right)_{div.}& = & (x_1^2+y_1^2) \left(\delta Z^{W_1}\right)_{div.}
+\frac{v_1^2}{(c_M^2 c_W^2)} \left( 
\delta Z^\gamma-2s_W^2\delta Z^{W_1}\right)_{div.}, \\ & & \nonumber \\
\left(\delta Z^{Z_2} \right)_{div.}& = & (x_2^2+y_2^2) \left(\delta Z^{W_2}\right)_{div.}
+\frac{v_2^2}{(c_M^2 c_W^2)} \left( 
\delta Z^\gamma-2s_W^2\delta Z^{W_2}\right)_{div.}.\end{eqnarray}
The last and certainly the hardest was to verify the finiteness of the renormalized $W_1$ vertex, which
would prove the finiteness of the muon decay amplitude. This has also been successfully done.

\section{Muon Decay at One-Loop}

We have to decompose all of the corrections into classic electromagnetic
and the weak ones. Let us remind, that the muon life time is described through the Fermi coupling
constant $G_F$, which, for historical reasons, does not contain electromagnetic radiative contributions 
to the Fermi theory.

\begin{figure}
\epsfig{file=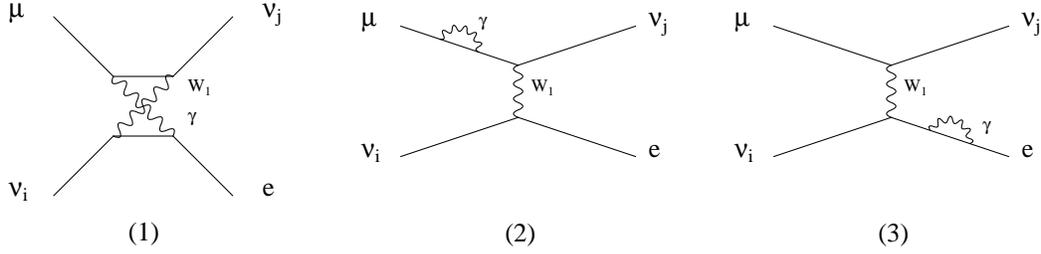, width=14cm}
\caption{Diagrams entering the electromagnetic corrections to the Fermi theory.}
\label{fermi-elec}
\end{figure}

The method of calculation has been devised by {\it Sirlin} \cite{eight}. He noted, that if we write the 
photon propagators in the muon and electron wave function, in the following way:
\begin{equation}
\frac{1}{k^2} = \frac{1}{k^2-M_{W_1}^2}-\frac{M_{W_1}^2}{k^2-M_{W_1}^2}\frac{1}{k^2},
\end{equation}
then diagrams of fig~\ref{fermi-elec} with the second term of the above equation, 
are, to an excellent approximation (of the order 
$\frac{\alpha}{4\pi}\frac{m_\mu^2}{M_{W_1}^2}$) equal to the electromagnetic corrections 
to the Fermi theory without bremsstrahlung.
One should, therefore, calculate radiative corrections to $G_F$ by simply neglecting these diagrams, but
using a massive photon on the external lepton lines\footnote{note however that the photonic corrections
to vertices and boson wave functions should contain a massless photon.}. This method has an additional
advantage, that we will not have to regularize the photonic corrections, since everything will be 
infrared finite.

Up to this point everything was the same as in the SM. One more point that is similar, is that we
can introduce a quantity, namely $\Delta r$, that will contain all the radiative corrections:
\begin{equation}
G_F = \frac{\pi\alpha}{\sqrt{2}}\frac{1}{M_{W_1}^2 s_W^2}\left(1+\Delta r\right).
\end{equation}
We can decompose $\Delta r$ into the following terms:
\begin{equation}
\Delta r = \frac{Re\Pi_{W_1 W_1}(M_{W_1}^2)-\Pi_{W_1 W_1}(0)}{M_{W_1}^2}-\Pi'_{\gamma\gamma}(0)
-\frac{\delta s_W^2}{s_W^2}+\delta_{V+B},
\label{delta_r}
\end{equation}
$\Pi'_{\gamma\gamma}$ comes from $\delta e$, and the rest of contributions is included
in $\delta_{V+B}$, which stands for vertex and box corrections.
 The first three are oblique corrections.

One might wonder, why we did not perform a Dyson resummation of the $W_1$ propagator. This is due to
the requirement of gauge invariance. Written in the above form the amplitude satisfies it, but if we
treat the bosonic propagator in a specific way, this gauge invariance will be lost. In SM model analyses, it
is usual to resum fermion loop contributions, since they can be shown to be gauge invariant. Here the
situation is more interesting. Apart from large fermion loops, also Higgs loops can be important. 
Therefore resumming fermion loops will lead to an asymmetry in the treatment of dominating effects. 
There is a way to solve the problem. We will not do it here, but to get better results one should
use methods of the type of pinch technique \cite{pinch} or background field \cite{back}. Further studies of 
these, applied to the LR model, will certainly be necessary\cite{pilo}.

\subsection{Top quark, Higgses  and Oblique Corrections}

The leading heavy particle contributions concentrate in $\delta s_W^2$.
Let us note that utmost importance
is attached to the top quark and Higgs in the SM. 

There  is a large difference
between the two models. In our case, we have two scales, one of them corresponds to the SM 
breaking
scale, but the second must be much larger. It so happens, that $\delta s_W^2$ corresponds 
to the second one.
Let us remind, that it contains only terms proportional to the ratio of mass renormalization 
constants and $v_R^2$ (see Eqs.(53,54)). 

In the Standard Model, the top quark came into play through the 
$\Delta \rho$ part of $\delta s_W^2$. 
Due to the existence of one scale, its contribution is, as we know \cite{hol2}:
\begin{equation}
(\Delta r)^{top}_{SM} = -\frac{c_W^2}{s_W^2}(\Delta\rho)^{top},
\end{equation}
where:
\begin{equation}
(\Delta\rho)^{top} = \frac{\sqrt{2}G_F}{16\pi^2}3m_t^2.
\end{equation}
In the LR model, the situation is much different, namely:
\begin{equation}
(\Delta r)^{top}_{LR} = \frac{\sqrt{2}G_F}{8\pi^2} c_W^2\left( \frac{c_W^2}{s_W^2}-1 \right)\frac{M_{W_1}^2}
{M_{W_2}^2-M_{W_1}^2}3m_t^2.
\end{equation}
For a $W_2$ boson mass of the order of $400$ GeV or larger, this contribution is smaller than the SM 
logarithmic
terms. This simply means, that with realistic masses of additional gauge bosons, we loose the
quadratic mass dependence of the oblique corrections on the top mass. Let us note, that one of the 
beautiful evidences for the correctness of the SM, came from the excellent agreement between the 
predicted top mass (from oblique corrections) and the actually measured at Fermilab. We have to acknowledge,
that the LR model cannot display this quality\footnote{This is a feature of all models with $\rho \neq 1$
as first noted by {\it Jegerlehner} \cite{jeg}.}. 

Let us consider the Higgses. In the SM there is custodial symmetry, which induces a cancellation of the
quadratic mass term of the Higgs. The remaining logarithms are weak enough not to allow for a precise
prediction of the still unknown Higgs mass.
Remark that quadratic mass terms put stringent bounds on the allowed masses. It could be feared, that
due to the lack of custodial symmetry, large Higgs masses, as required by $FCNC$ bounds, would not
be allowed by the oblique corrections. The LR model in this minimal version would be ruled out. 
Fortunately for us, these terms occur in ratios with the large symmetry breaking scale. Therefore, the
risk is lessened. 
We can give a typical Higgs contribution. This one is for the lightest, supposedly analogue of
the SM Higgs:
$$
(\Delta r)^{\mbox{\it lightest Higgs}}_{LR} = \frac{\sqrt{2}G_F}{48\pi^2}\left(\frac{M_{W_1}^2}{M_{W_2}^2}
\frac{c_W^2}{s_W^2}(1-2s_W^2)+
\frac{M_{W_1}^2}{M_{Z_2}^2}\frac{1}{s_W^2}(4c_W^2-1)\right)M_{H^0_0}^2.
$$
\begin{equation}
\end{equation}
The remaining Higgses give terms of comparable structure.

\subsection{Heavy Neutrinos}

Contrary to the top, heavy neutrinos give important contributions proportional to squares of the masses.
We have:
\begin{equation}
(\Delta r)^{N}_{LR} = \sum_{N=heavy} \frac{\sqrt{2}G_F}{16\pi^2}c_M^2 c_W^2 
\frac{M_{W_1}^2}{M_{W_2}^2-M_{W_1}^2}m_N^2.
\end{equation}

The exact contribution depends heavily on the relation between neutrino
and $W_2$ masses. If we assume (three) heavy neutrinos, to be of the order
of $M_{W_2}$, then from Eqs.(63,67) we have $(m_N \simeq M_{W_2}
>>M_{W_1})$

\begin{equation}
| (\Delta r)^{N}_{LR}/(\Delta r)^{top}_{SM} | \simeq c_M^2 s_W^2
\frac{M_{W_1}^2}{m_t^2} \simeq 0.02,
\end{equation}

so the heavy neutrino contribution of oblique corrections to $\Delta r$ is
smaller
than the SM top one. However, already with $m_N \simeq 7 M_{W_2}$ this
is comparable\footnote{Let us note however, that such heavy neutrinos are
at the edge of perturbation theory where (assuming neutrino Yukawa
couplings to be smaller than one), a natural limit can be derived,
$m_N \leq 2M_{W_2}/g$ \cite{gluzainv}}.
Interestingly, neutrino contributions come with the opposite sign to the SM top quark one,
similarly as in \cite{bert}.

Finally let us comment shortly on heavy neutrino contributions to vertices and boxes.
In the SM, vertex corrections were not negligible, but  small and under good control.
The reason was, that as mass terms of the leptons were unimportant, they could be parametrized as
functions of the gauge boson masses. Moreover, Higgs diagrams could be neglected altogether.
The situation changes drastically in the LR model. 
The corrections now depend seriously not only on masses, but also on the mixing matrices. 
The largest terms, generated by Higgs diagrams, can in principle exhibit 
a quadratic dependence on the
neutrino masses. This can be easily seen, with some dimensional analysis. Fortunately, 
it has been 
proved in the case of the SM with additional right handed singlets \cite{melo}, that these terms cancel with
similar contributions from the external wave functions. Since the proof can be made by 
explicit calculation\footnote{and does not depend on anything more than just properties of one-loop
scalar functions.}, and in this respect nothing changes if we go over to the LR model, here also
these terms cancel. 
Remain, however, contributions proportional to logarithms of ratios of neutrino masses to light gauge boson
masses. They are not negligible, but to make any predictions, concrete mixing matrices and masses must
be considered \cite{fut}.

The situation is more complex with boxes. 
This time, there is nothing that contributions proportional to squares of neutrino masses could cancel
against. It can be conjectured, that these terms will put stringent bounds on allowed neutrino masses and
mixings. 

\section{Conclusions and Outlook}

We have performed a full one loop calculation of a physical process, the muon decay,
in the framework of the left-right symmetric model. The calculation involved
a huge number of diagrams. It has thus been performed using the computer program FORM, and
custom programs.
 
A simple renormalization scheme has been developed for four-fermion processes. It has been 
shown to work, by a direct check of divergence cancellation. Its relation to the Standard Model
results has been studied.

Later, contributions of heavy particles through oblique corrections have been discussed. 
It has been
shown that the strong dependence on the top quark mass, so famous in the SM, has been lost. The size
of Higgs and heavy neutrino diagrams has been demonstrated
 not to exceed reasonable bounds, thereof leading
to the conclusion that the model is not trivially ruled out.

At last, we have to note, that $\Delta r$ of the LR model is much different, apart from the $\Delta\alpha$
contributions, from its SM counterpart. It means, that if one makes bounds from tree level diagrams, corrected
by SM $\Delta r$, the result is only a rough approximation of reality. 
In fact as seen in the last section, in
principle $\Delta r$ can be anything.

As an outlook of the future, we have to state that many issues remain to be considered. First and foremost,
fits to low- and high-energy processes are now possible \cite{fut}. 
A lot of effort is nevertheless required on the phenomenological side. Without wise guesses on the possible
heavy neutrino sector, it would be impossible to derive any numerics. Thus, it again shows, that 
light states are nontrivially entangled with unobservable heavy ones.

\section*{Acknowledgements}
This work was supported by Polish Committee for Scientific Researches under
Grant No. 2P03B08414 and 2P03B04215. J.G. would like to thank F. Jegerlehner 
for valuable remarks
and Alexander von Humboldt-Stiftung for fellowship.


\begin{thebibliography}{99}
\bibitem{hol1} W. Hollik, Z. Phys. {\bf C37} (1988) 569.
\bibitem{hol2} W. Hollik, {\it Talk given at the Cracow Epiphany
Conference, Cracow, Poland, 9-11 Jan. 1999.}
\bibitem{melo} P. Kalyniak, I. Melo, Phys.Rev. {\bf D55} (1997) 1453.
\bibitem{non} T. Appelquist, J. Carazzone, Phys. Rev. {\bf D11} (1975)
2856.
\bibitem{senj} G. Senjanovic, A. Sokorac,
Phys. Lett. {\bf B76} (1978) 610; Phys. Rev. {\bf D18} (1978) 2708.
\bibitem{sup}  Super-Kamiokande Collaboration, (Y. Fukuda {\em et al.}),
Phys. Rev. Lett. {\bf 81}  (1998) 1562.
\bibitem{jez} T. Yanagida, {\em Proc. of the Workshop
on Unified Theory and Baryon Number in the Universe},
eds. O. Sawada and A. Sugamoto (KEK, 79-18, 1979);

M. Gell-Mann, P. Ramond, and R. Slansky, in {\em Supergravity},
ed. by
P. van Neiuwenhuizen and D. Freedman (North-Holland, Amsterdam, 1979);

M. Je\.zabek and Y. Sumino,  hep-ph/9904382 and references therein.

\bibitem{box}   G. Beall, M. Bander and A. Soni, Phys. Rev. Lett. {\bf 48}
(1982) 848.

\bibitem{pilo} Z. Gagyi-Palffy, A. Pilaftsis, K. Schilcher, Nucl.Phys. {\bf B513} (1998) 517.

\bibitem{fut} M. Czakon, J. Gluza and M. Zra\l ek, in progress.

\bibitem{obliq} M. E. Peskin, T. Takeuchi, Phys. Rev. {\bf D46} (1992) 381;
 D.C. Kennedy, B.W. Lynn, Nucl. Phys. {\bf B322} (1989) 1;
K. Hagiwara, D. Haidt, S. Matsumoto, Eur. Phys. J. {\bf C2} (1998) 95. 
\bibitem{cust} M. Veltman Acta Phys. Pol. {\bf B8} (1977) 475.
\bibitem{lr}  J.C. Pati and A. Salam, Phys. Rev. {\bf D10} (1974) 275;

R.N. Mohapatra, J.C. Pati, Phys. Rev. {\bf D11} (1975) 2558;

 R.N. Mohapatra, J.C. Pati, ibid. {\bf D11} (1975) 566; 

 G. Senjanovic and R.N. Mohapatra, ibid. {\bf D12} (1975) 1502; 

G. Senjanovic, Nucl. Phys. {\bf B153} (1979) 334.

I. Liede, J. Maalampi and M. Ross Nucl. Phys. {\bf B146} (1978) 157;

V. Barger, E. Ma and K. Whisnant, Phys Rev {\bf D26};
(1982) 2378;

T.Rizzo and G Senjanovic Phys. Rev. {\bf D24} (1981) 704;

X. Li and D. Marshak, Phys. Rev.{\bf D25} (1982) 1886;

J.E. Kim, P. Langacker, M. Levine and H. Wiliams, Rev. Mod. Phys.
{\bf 53} (1981) 211;

R.N. Mohapatra in 'CP violation' ed. by C. Jarlskog, 1988;

W. Grimus {\em Lecture given at the 4th Hellenic School on elementary Particle Physics,
Corfu, 2-20 September 1992}.

\bibitem{gun1}   J.F. Gunion, J. Grifols, A. Mendez, B. Kayser, F. Olness,
Phys. Rev. {\bf D40} (1989) 1546.

\bibitem{gun2} N.G. Deshpande, J.F. Gunion, B. Kayser, F. Olness,
Phys. Rev. {\bf D44} (1991) 837.
\bibitem{glu} J. Gluza, M. Zralek, Phys. Rev. {\bf D51} (1995) 4695. 

\bibitem{reszta}  R.N. Mohapatra, P.B. Pal, Phys. Rev. {\bf D38} (1988) 2226;

B.S. Balakrishna and R.N. Mohapatra, Phys. Lett. B216 (1989) 349;

S. Rajport, Phys. Lett. B191 (1987) 122; 

A. Davidson and K.C. Wali, Phys. Rev. Lett. 59 (1987) 393; 

B.S. Balakrishna, Phys. Lett. 60 (1988) 1602; 

K.S. Babu and R.N. Mohapatra, Phys. Rev. Lett. 62 (1989) 1079;

R.N. Mohapatra, Phys. Lett. B201 (1988) 517; 

B.S. Balakrishna, A. Kagan and R.N. Mohapatra, Phys. Lett. B205 (1988) 345; 

B.S. Balakrishna, E. Ma Phys. Lett. B214 (1988) 267; 
Phys. Rev. Lett. 63 (1989) 349.
\bibitem{fcnc} G. Ecker et al., Phys. Lett. {\bf B94} (1980) 381;
Nucl. Phys. {\bf B250} (1985) 517.
\bibitem{glu1} J. Gluza, M. Zra\l ek, Phys. Rev. {\bf D48}
(1993) 5093. 
\bibitem{jap} M. Doi, T. Kotani, E. Takasugi, Prog. Theor. Phys. {\bf 71}
(1984) 1440.

\bibitem{eight} A. Sirlin, Phys. Rev. {\bf D22} (1980) 971.
\bibitem{hol} W. Hollik, Fort. Phys. {\bf 38} (1990) 165.
\bibitem{sack}  A.Denner and T. Sack, Nucl. Phys. {\bf B347} (1990) 203.
\bibitem{kniehl}  B. Kniehl,  A. Pilaftsis, Nucl. Phys. {\bf B474} (1996) 286.
\bibitem{duka} P. Duka, {\it 'Quantization and renormalization of the
left-right symmetric models'}, Ph. D. Thesis, Katowice, 1999.
\bibitem{verm} 'Symbolic manipulations with FORM', J.A.M. Vermaseren.
\bibitem{pinch} J. Papavassiliou, Phys. Rev {\bf D41} (1990) 3179,

G. Degrassi, A. Sirlin, Phys. Rev. {\bf D46} (1992) 3104, and references therein.
\bibitem{back} L. F. Abbott, Nucl. Phys. {\bf B185} (1981) 189,
Acta. Phys. Pol. {\bf B13} (1982) 33, and references therein.

\bibitem{jeg} F. Jegerlehner, Prog. Part. Nucl. Phys. {\bf 27} (1991) 1. 
\bibitem{gluzainv} J. Gluza, M. Zralek, Phys. Rev. {\bf D52} (1995) 6238. 
\bibitem{bert} S. Bertolini, A. Sirlin, Phys. Lett. {bf 257} (1991) 179.
     
\end{thebibliography}
\end{document}